\documentclass[acmsmall]{acmart}
\usepackage{booktabs}
\usepackage{enumitem} 
\usepackage{multirow}
\usepackage{array}
\usepackage[table]{xcolor}
\usepackage{graphicx}
\usepackage{tabularx}
\usepackage{wrapfig}             
\usepackage{color}  
\usepackage{mdframed}
\usepackage{pifont}
\usepackage{makecell}
\usepackage{booktabs}
\usepackage[most]{tcolorbox}
\usepackage{xspace}

\AtBeginDocument{%
  }

\setcopyright{cc}
\setcctype{by}
\acmDOI{10.1145/3832118}
\acmYear{2026}
\acmJournal{PACMSE}
\acmVolume{3}
\acmNumber{ISSTA}
\acmArticle{ISSTA027}
\acmMonth{10}
\acmSubmissionID{issta26main-p273-p}
\received{2026-01-30}
\received[accepted]{2026-04-16}

\begin{document}

\title{PackMonitor: Towards Zero Package Hallucinations via Decoding-Time Monitoring}

\author{Xiting Liu}
\orcid{0009-0002-2896-0073}
\affiliation{%
  \institution{Tsinghua University}
  \department{Department of Computer Science and Technology}
  \city{Beijing}
  \country{China}
}
\email{liuxt25@mails.tsinghua.edu.cn}

\author{Yuetong Liu}
\orcid{0009-0003-6512-2412}
\affiliation{%
  \institution{Tsinghua University}
  \department{College of AI}
  \city{Beijing}
  \country{China}
}
\email{23371522@buaa.edu.cn}

\author{Yitong Zhang}
\orcid{0009-0000-1138-4503}
\affiliation{%
  \institution{Tsinghua University}
  \department{College of AI}
  \city{Beijing}
  \country{China}
}
\email{22373337@buaa.edu.cn}

\author{Jia Li}
\orcid{0000-0002-5579-8852}
\affiliation{%
  \institution{Tsinghua University}
  \department{College of AI}
  \city{Beijing}
  \country{China}
}
\email{jia\_li@mail.tsinghua.edu.cn}

\author{Shi-Min Hu}
\orcid{0000-0001-7507-6542}
\affiliation{%
  \institution{Tsinghua University}
  \department{Department of Computer Science and Technology}
  \city{Beijing}
  \country{China}
}
\email{shimin@tsinghua.edu.cn}

\renewcommand{\shortauthors}{Xiting Liu, Yuetong Liu, Yitong Zhang, Jia Li, and Shi-Min Hu}
\newcommand{\name}{\textit{\textsc{PackMonitor}}\xspace}

\begin{abstract}
  As Large Language Models (LLMs) are increasingly integrated into software development workflows, their trustworthiness has become a critical concern. In dependency recommendation scenarios, the reliability of LLMs is undermined by widespread package hallucinations, where models often recommend hallucinated packages. Recent studies have proposed a range of approaches to mitigate this issue. Nevertheless, existing approaches typically merely reduce hallucination rates rather than eliminate them, leaving persistent software security risks.

In this work, we argue that package hallucinations are theoretically preventable based on the key insight that package validity is decidable through finite and enumerable authoritative package lists. Building on this, we propose \name, the first approach capable of fundamentally eliminating package hallucinations by continuously monitoring the model's decoding process and intervening when necessary. To implement this in practice, \name addresses three key challenges: 
\ding{182} determining when to trigger intervention via a Context-Aware Parser that continuously monitors model outputs and selectively activates intervening only during installation command generation; 
\ding{183} resolving how to intervene by employing a Package-Name Intervenor that strictly limits the decoding space to an authoritative package list; and 
\ding{184} ensuring monitoring efficiency through a DFA-Caching Mechanism that enables scalability to millions of packages with negligible overhead. 
Extensive experiments on five widely used LLMs demonstrate that \name is a training-free, plug-and-play solution that consistently reduces package hallucination rates to zero while maintaining low-latency inference and preserving original model capabilities.
\end{abstract}

\ccsdesc[500]{Security and privacy~Software security engineering}
\ccsdesc[500]{Computing methodologies~Natural language processing}
\ccsdesc[300]{Theory of computation~Grammars and context-free languages}

\keywords{Package Hallucination, Large Language Models, Constrained Decoding}

\maketitle
\section{Introduction}
\label{1}
In recent years, Large Language Models (LLMs) such as Qwen2.5-Coder~\cite{Qwen2.5-Coder} and DeepSeek-Coder~\cite{DeepSeekCoder}, along with LLM-powered agents like Copilot~\cite{github_copilot} and Cursor~\cite{cursor}, have been reshaping the landscape of software engineering~\cite{intro_first, cai2025ai, yang2025difftester, li2025beyond, li2025structured, zhang2026see}. As these systems become deeply integrated into real-world software development workflows, their trustworthiness has become a critical concern~\cite{ai_programming, codehallucination2, zhang2026lookahead, zhang2026grammar}. Unfortunately, current models frequently exhibit hallucinations~\cite{CodeHallucation, codehallucination2, codehallucination3}, which substantially undermine trustworthiness and introduce risks to the software lifecycle.

\begin{figure}[t!]
    \centering
    \includegraphics[width=0.96\textwidth]{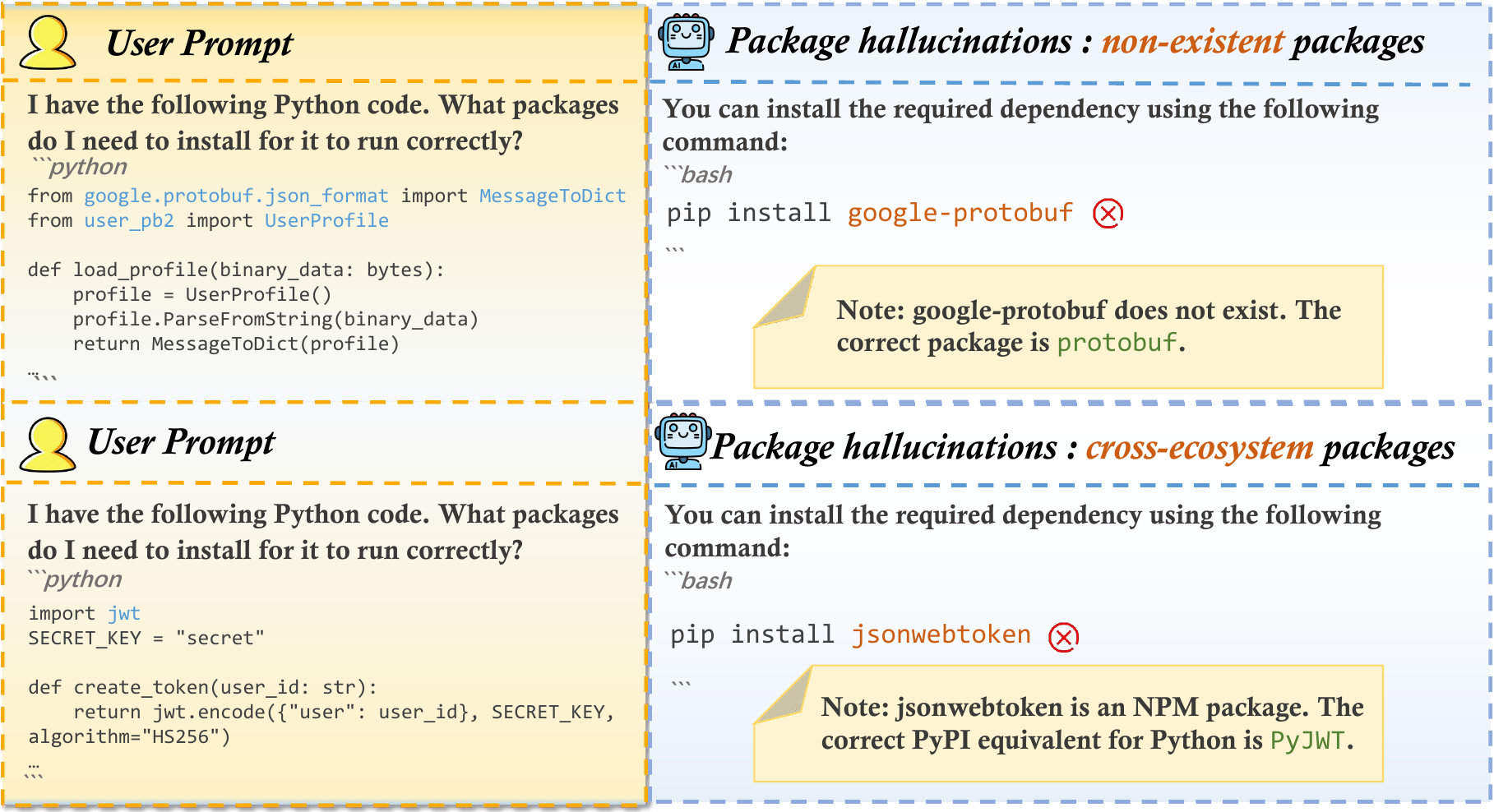}
    \vspace{-3pt}
    \caption{Examples of active and passive deception in software engineering.}
    \label{fig:intro-case}
    \vspace{-0.18in}
\end{figure}

Among the various forms of hallucination, package hallucination has received particular attention~\cite{weHaveAPackageForYou, pacakge_hallucination1, pacakge_hallucination2}, where a model may recommend non-existent software packages (top of Figure~\ref{fig:intro-case}) or suggest packages from the wrong ecosystem (bottom of Figure~\ref{fig:intro-case}). 
Prior work shows that package hallucinations are widespread: across 16 widely used LLMs, hallucinated packages can account for up to 19.7\% of recommended packages~\cite{HFuzzer}, substantially undermining model trustworthiness.
We argue that widespread package hallucinations create a concrete attack surface for software supply chain attacks~\cite{supply_chain_attack}. In practice, LLMs often repeatedly output a small set of hallucinated package names, enabling adversaries to preemptively register them as malicious packages. When such a name is later recommended and mistakenly installed, the malicious dependency can be introduced into the software stack, triggering downstream compromises.

The severity of package hallucinations has motivated a range of approaches aimed at mitigating them~\cite{hallucination-mitigation, rag, self-relection, sft}. However, we find that existing approaches can only reduce the likelihood of package hallucinations, rather than guarantee their elimination. For example, Retrieval-Augmented Generation (RAG) injects retrieved valid package names into the prompt, yet on DeepSeek-Coder-6.7B~\cite{DeepSeekCoder} it only reduces the hallucination rate from 16.14\% to 12.24\%. Self-Refinement (SR) asks the model to verify and regenerate its own outputs, but yields only marginal gains (e.g., CodeLlama~\cite{codellama} drops from 26.28\% to 25.51\%). Supervised fine-tuning (SFT) retrains models on hallucination-free data, but still cannot fully eliminate hallucinations and may also degrade utility. As a result, the remaining package hallucinations continue to pose an exploitable software supply chain threat~\cite{supply_chain_attack, package_confusion2}, hindering the broader adoption of LLMs in software development.

To improve the trustworthiness of LLMs, we propose \name, the first approach capable of fundamentally eliminating package hallucinations. Our key insight is that valid package names reside within a finite, enumerable authoritative package list (e.g., the PyPI index~\cite{pypi})~\cite{weHaveAPackageForYou}. This makes package-name validity decidable, rendering package hallucinations theoretically preventable. Building on this insight, we propose \name, which continuously monitoring the model's decoding process and intervening when necessary.
Concretely, it intervenes in the decoding phase by pruning any generation path that would inevitably lead to a non-existent package name via logits masking. In this way, hallucinated packages are excluded from the generation space, guaranteeing a zero package-hallucination rate. Notably, \name requires no additional training and operates as a lightweight, plug-and-play solution.

However, implementing \name in practice comes with three key challenges.
\ding{182} First, \name must determine \textbf{when to trigger intervention}. 
This is crucial because security risks stem exclusively from executable package installation commands. Applying intervention globally would inevitably penalize benign text or ordinary code, thereby reducing the model's utility.
To address this, \name employs a \textbf{Context-Aware Parser} that operates in a continuous monitoring mode but activates strict intervention only when detecting package-name generation following installation commands. This selective mechanism ensures that the model's original generation capabilities remain preserved outside the high-risk scope.
\ding{183} Second, we address \textbf{how to intervene LLMs' generation}. 
To guarantee zero package hallucinations under the enforced package list, \name employs a \textbf{Package-Name Intervenor}. Upon activation, the intervenor leverages a Deterministic Finite Automaton (DFA) constructed from an authoritative package list to precisely delimit the valid generation space.
At each decoding step, the intervenor derives the exact set of permissible next tokens and rules out any illegal continuations. Consequently, any path leading to a hallucinated package name is strictly removed from the generation space.
\ding{184} Third, we address \textbf{how to perform monitoring efficiently}. In practice, the scale of valid packages is massive—for instance, PyPI contains over 700k packages. At this scale, the resulting DFA becomes extremely large, making on-the-fly construction prohibitively slow. To ensure scalability, \name introduces a \textbf{DFA-Caching Mechanism}: it pre-constructs the DFA into a persistent checkpoint and loads it into memory during inference. This "build once, reuse many" strategy decouples construction costs from runtime execution, enabling low-latency serving even for massive ecosystems.

We conducted extensive experiments on two benchmarks, including HFuzzer~\cite{HFuzzer} and Package4U~\cite{weHaveAPackageForYou} for evaluating package hallucinations, using five widely used LLMs and multiple baselines. Our findings are as follows: 
\ding{182} Across all experimental settings, \name strictly reduces package hallucination rates to zero. For instance, it successfully eliminate the 11.6\% package hallucination rates originally observed in DeepSeek-Coder-6.7B.
\ding{183} \name introduces only negligible inference overhead, with the generation time per response increasing by merely 0.05–0.3 seconds.
\ding{184} Unlike SFT, which can significantly degrade code generation capability, our results on HumanEval~\cite{humaneval} indicate that equipping models with \name does not harm their utility.

In summary, our contributions are threefold:
\begin{itemize}[nosep]
    \item We prove that package hallucination is a unique type of factuality hallucination that can be theoretically prevented, owing to its deterministic nature, which allows candidate package names to be verified against finite authoritative package lists.
    \item We propose \name, the first approach capable of completely eliminating package hallucinations. \name integrates a context-aware parser with a package-name intervenor and an efficient DFA-caching mechanism.
    \item We conduct extensive experiments using two benchmarks on five widely used LLMs. The results show that \name reduces package hallucination rates to zero across all settings while maintaining low-latency inference.
\end{itemize}

\vspace{-4pt}
\section{Related Work and Motivation}
In this section, we present the essential background for understanding our proposed approach and discuss the research motivation behind it.

\vspace{-3pt}
\subsection{LLM Hallucinations and Package Hallucinations} 
\label{2.1}
With the rapid development of Large Language Models (LLMs), \textbf{LLM hallucinations}~\cite{hallucination_survey,hallucination_survey2,hallucination_survey3} have become a widely recognized issue, where LLMs generate misleading, unsubstantiated, or completely fictitious information~\cite{codehallucination2}. LLM hallucinations can be broadly classified into \textbf{factuality hallucinations} and \textbf{faithfulness hallucinations}~\cite{faithfulness_and_factuality,metamorphic_testing}. Factuality hallucinations arise when the generated content deviates from verifiable real-world facts, whereas faithfulness hallucinations occur when there are inconsistencies between the generated content and the given input context.~\cite{hallucination_survey}.

In recent years, LLM-powered coding agents have garnered considerable attention with their rapid development~\cite{ai_programming, openhands, chen2025featbench}, highlighting the immense potential of LLMs in software development. However, the impact of hallucinations in this domain has been further magnified. \textbf{Package hallucination}~\cite{weHaveAPackageForYou, pacakge_hallucination1, pacakge_hallucination2} is a specialized form of factuality hallucination, in which LLMs generate non-existent or ecosystem-incompatible package names in the output.~\citeauthor{weHaveAPackageForYou}~\cite{weHaveAPackageForYou} conducted an evaluation on 16 open-source and closed-source LLMs, generating over 1 million package recommendations from around 40,000 programming questions collected from LLM-generated content and Stack Overflow~\cite{stackoverflow}. Their findings indicated that 19.7\% of the recommended packages were hallucinatory, with 205,474 distinct non-existent package names identified. In addition,~\citeauthor{HFuzzer}~\cite{HFuzzer} proposed HFuzzer, a fuzzing-based framework to evaluate whether models generate hallucinated packages. The experiments using HFuzzer demonstrate that most models, including both open-source and proprietary commercial models, produce a substantial number of hallucinated packages.

In automated coding agent workflows, invalid package dependencies may cause build failures and deployment disruptions, thereby increasing the overhead of development and debugging processes~\cite{auto-build-env}. 
More dangerously, as discussed in Section~\ref{1}, package hallucinations introduce a critical attack vector for \textbf{software supply chain attacks}~\cite{supply_chain_attack,package_confusion,package_confusion2}.
Attackers no longer need to guess or obfuscate popular package names; instead, they can monitor the outputs of mainstream LLMs at scale and collect the frequently generated hallucinated package names for targeted preemptive registration.
Once an automated coding agent installs such a maliciously registered package, the attacker’s code may be executed within the agent’s runtime environment, leading to severe consequences such as data exfiltration, unauthorized network access, and further compromise of downstream systems in the software development pipeline~\cite{supply_chain_attack,package_confusion,package_confusion2} .

\subsection{Mitigation of Package Hallucinations} 
\label{2.2}
Existing methods for mitigating package hallucinations in LLMs can be broadly categorized by whether they require additional model training, which yields two primary categories: \textbf{training-free} and \textbf{training-based} methods~\cite{hallucination-mitigation, weHaveAPackageForYou}. The former mitigates hallucinations via prompt-based guidance or post-hoc verification, while the latter relies on parameter-level adjustments to the model.

\vspace{3pt} \noindent
\textbf{Training-free Methods.} To avoid the computational costs and unintended side effects of retraining, several training-free methods have been proposed to mitigate package hallucinations without modifying model parameters. Two representative techniques are \textbf{Retrieval-Augmented Generation (RAG)}~\cite{rag, rag2} and \textbf{Self-Refinement (SR)}~\cite{self-refinement,self-relection}. RAG mitigates package hallucinations by injecting externally retrieved, valid package information into the prompt, thus enhancing factual grounding during generation. Although this approach can reduce hallucination rates in certain scenarios, its effectiveness is often limited and can vary across settings. Empirical findings demonstrate that RAG only yields modest performance improvements; for example, it reduces the package hallucination rate of DeepSeek-Coder-6.7B from 16.14\% to 12.24\%~\cite{weHaveAPackageForYou}. SR, on the other hand, prompts the LLM to verify and optionally regenerate content following the initial generation step. This method frames hallucination mitigation as a post-hoc correction task, which relies on the model’s ability to identify its own factual errors. Although SR can achieve marginal reductions in package hallucination rates, it is inherently unreliable: once an LLM generates hallucinated packages with high confidence, SR often fails to detect or correct such errors. Additionally, SR incurs extra inference overhead, as it requires additional verification and regeneration steps during the inference process. Recent work has also shown that decoding-time monitoring can improve the
reliability of code generation. ~\citeauthor{agrawal2023monitor}use static analysis over repository context to guide Code LM decoding and improve compilation-related correctness properties~\cite{agrawal2023monitor}. Our work is complementary: instead of enforcing repository-level type or API constraints, we focus on preventing invalid package names in generated installation commands.

\vspace{3pt} \noindent
\textbf{Training-based Methods.} Several studies have mitigated package hallucinations via \textbf{Supervised Fine-Tuning (SFT)}~\cite{sft, weHaveAPackageForYou}. In this paradigm, LLMs are fine-tuned on carefully curated datasets that exclude hallucinated package names.By adjusting the model’s parameters, SFT can significantly reduce the model’s tendency to generate hallucinated package names. For example,~\citeauthor{weHaveAPackageForYou}~\cite{weHaveAPackageForYou} demonstrate that SFT can reduce the package hallucination rates to as low as 2.66\% for certain LLMs. However, training-based mitigation suffers from several fundamental limitations. First, SFT requires the construction of high-quality, hallucination-free training datasets, which is expensive and difficult to maintain as software ecosystems evolve. Second, altering the model’s parameters may introduce unintended side effects; previous studies have shown that SFT can cause a significant degradation in the model’s general code generation performance\cite{weHaveAPackageForYou}. Most importantly, despite its effectiveness in reducing package hallucination rates, SFT does not fully eliminate package hallucinations under existing settings. Consequently, package hallucination remains possible, which leaves residual security vulnerabilities in the downstream software supply chain.

\vspace{3pt} \noindent
\textbf{Summary.} Despite their inherent differences, both training-free and training-based methods suffer from a fundamental limitation: they can only mitigate package hallucinations rather than completely eliminate them. Training-free methods rely on prompt-based guidance or post-hoc verification, thus offering no strict generation-time guarantees; training-based methods reduce package hallucinations at the expense of retraining overhead and degraded model performance. As long as the package hallucination rate remains non-zero, such hallucinations can still introduce exploitable security vulnerabilities, which motivates the need for novel methods that can fundamentally prevent the generation of hallucinated packages.

\subsection{Constrained Decoding for Code Generation}

Constrained decoding~\cite{constrained_decoding_1, constrained_decoding_2, constrained_decoding_3} has been explored as a way to improve the reliability and controllability of LLM generation. Instead of relying solely on post-hoc verification or repair, constrained decoding incorporates external constraints directly into the decoding process, pruning invalid generation paths before they are sampled. In code generation, such constraints are particularly useful because many correctness properties can be partially specified by formal structures, such as grammars, type systems, static-analysis rules, API specifications, or repository-level program context.

Recent studies have applied constrained decoding to different software engineering tasks. For example, grammar-based or syntax-aware decoding can ensure that generated code follows certain syntactic structures~\cite{melcer2024constrained}. Other approaches incorporate semantic constraints, such as type constraints or API-usage constraints, to guide model generation toward code that is more likely to compile or match the surrounding program context~\cite{wang2024fantastic, fu2024constrained}. Closely related to our work, ~\citeauthor{agrawal2023monitor} propose monitor-guided decoding, which uses static analysis over repository context to guide Code LM decoding and avoid invalid continuations, such as unavailable or inconsistent API usages~\cite{agrawal2023monitor}. These studies show that decoding-time constraints can effectively improve code generation reliability by leveraging deterministic information from programming languages and software repositories.

Our work is inspired by this line of research, but focuses on package hallucination, where the validity of generated package names can be checked against a finite and enumerable package registry. We further discuss how this observation motivates the design of \name in the next section.

\subsection{Motivation}

The constrained decoding perspective provides a natural way to revisit package hallucinations. Package installation in software development exhibits a key characteristic that distinguishes it from many open-ended factuality problems: package-name validity is decidable under a given software ecosystem. For a given package ecosystem, such as PyPI~\cite{pypi}, npm~\cite{npm}, or Maven~\cite{maven}, a finite and enumerable list of valid package names can be obtained as ground truth, either from an official registry snapshot or a curated allowlist. Thus, for any candidate package name, its membership in this list is unambiguous: it either belongs to the set or not.

This decidability property allows package hallucination to be formulated as a constrained decoding problem. Consider the hallucinated package example in Figure~\ref{fig:intro-case}: when prompted to install Protocol Buffers, an LLM may suggest “google-protobuf” due to spurious statistical correlations learned from its training data. While “google.protobuf” is a legitimate Python import path, it is not a valid package name in the predefined authoritative package list. With access to such an allowlist, \name intervenes in real time as the LLM generates a package name token by token. Conceptually, \name intervenes as follows: for each candidate next token, \name checks whether appending the token would keep the current partial string consistent with at least one allowlisted package name (i.e., whether the updated prefix remains a valid prefix for any legitimate package name in the allowlist). If a token would lead the generation process to a state where no allowlisted package name can be matched by any subsequent continuation, selecting the token would inevitably result in an invalid package name; \name thus masks the token by setting its corresponding logit value to $-\infty$. Meanwhile, a termination token (e.g., whitespace or newline) is allowed only if the current prefix already constitutes a complete, allowlisted package name. In this manner, \name prunes all token choices that would inevitably result in a hallucinated package, while preserving all valid package name completions.

However, making this idea practical requires addressing two challenges that are specific to package hallucination. First, package-name constraints should not be applied globally. LLM responses often contain natural language explanations, code snippets, comments, and ordinary shell commands. Enforcing package-name constraints over the entire output would unnecessarily restrict the model and may degrade its general generation capability. Therefore, \name introduces a context-aware parser that continuously monitors the output but activates strict constrained decoding only when package names are being generated in installation commands. Second, package ecosystems are large and continuously evolving, making naive constraint construction inefficient. To address this issue, \name compiles the authoritative package list into a DFA and introduces a caching mechanism to reuse the constructed automaton during inference. In this way, \name adapts the general idea of constrained decoding to the package hallucination problem by formalizing package-name validity as a finite constraint space, applying constraints only in the appropriate generation context, and making the approach efficient enough for large-scale package registries.

\section{Methodology}

In this section, we present the design of \name, a decoding-time monitoring framework that is capable of eliminating package hallucination. We begin with a high-level overview of the framework in Section~\ref{3.1}. We then detail the precise timing of intervention in Section~\ref{3.2}, the mechanics of the intervention process in Section~\ref{3.3}, and the optimization strategies for efficient monitoring in Section~\ref{3.4}.

\begin{figure}[t!]
    \centering
    \includegraphics[width=0.98\textwidth]{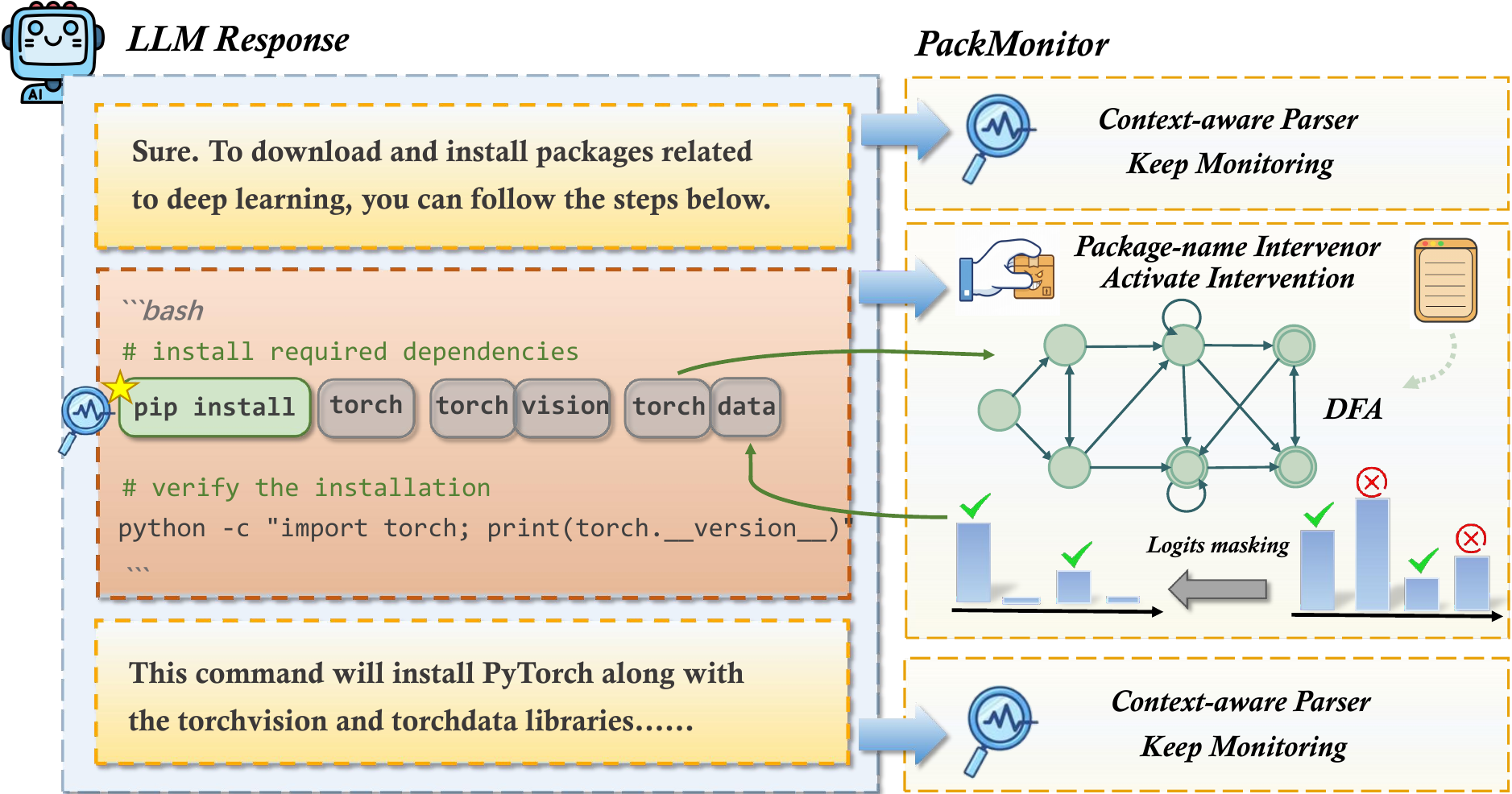}
    \vspace{-3pt}
    \caption{Overview of our proposed \name{}.}
    \label{fig:overview}
    \vspace{-0.18in}
\end{figure}

\subsection{Overview}
\label{3.1}

Driven by the insight that package validity is a deterministically decidable problem within a finite and enumerable scope, we propose \name. The framework orchestrates two tightly coupled components that operate synchronously with LLM inference: the \textbf{Context-Aware Parser} and the \textbf{Package-Name Intervenor}.

Figure~\ref{fig:overview} illustrates the overall framework of \name. 
First, the context-aware parser operates in a continuous monitoring mode to precisely locate the generation of package names following installation commands. 
Upon detecting such a context, it triggers the \textbf{package-name intervenor}. 
This module leverages a Deterministic Finite Automaton (DFA), constructed from authoritative package lists, to perform token-level validity verification. It enforces strict compliance by dynamically applying logits masks, thereby blocking any token that would lead to invalid package names.

To tackle the scalability challenge posed by modern ecosystems containing millions of packages, we further introduce a \textbf{DFA-Caching Mechanism}. This mechanism decouples the computationally intensive DFA construction from the runtime inference. By pre-compiling the DFA offline and loading the checkpoint during inference, \name effectively amortizes the initialization cost, enabling it to scale robustly to massive ecosystems with negligible latency overhead.

\newcommand{\softmax}{\mathrm{softmax}}

\subsection{When to Trigger Intervention?}
\label{3.2}

In this section, we address the first challenge: determining the precise timing for intervention. A naive approach that intervenes the entire output stream against a package list is infeasible; it would inevitably penalize benign natural language tokens or generic code, thereby severely degrading the model's general utility. We argue that supply chain security risks stem exclusively from executable package installation commands. Therefore, while monitoring must be continuous, strict intervention must be confined to these high-risk regions to ensure that the model's original generation capability remains untouched.

To achieve this, \name employs a \textbf{context-aware parser} modeled on Context-Free Grammar (CFG). This parser functions as a real-time sentinel, continuously monitoring generated tokens to distinguish between risk-free contexts and package-name generation zones.

Formally, we model the response of an LLM as an interleaving sequence of natural language and code segments:
\begin{equation}
\texttt{Response} \rightarrow (\texttt{NATURAL\_LANGUAGE} \mid \texttt{CODE})^{*}.
\end{equation}
Code blocks are identified by standard delimiters (e.g., \verb|```bash| and \verb|```|). Within these code blocks, the parser further differentiates between actual installation commands—which constitute the attack surface that we focus on—and ordinary code lines (such as comments or variable assignments):
\begin{equation}
\texttt{CODE} \rightarrow (\texttt{INSTALL\_CMD\_LINE} \mid \texttt{ORDINARY\_CODE\_LINE})^{*}.
\end{equation}
While we use the prevalent \texttt{pip install} command for PyPI packages as a running example, our approach is general and can be instantiated for installation commands in other ecosystems (discussed in Section~\ref{6.2}). Under the PyPI ecosystems, an installation command line is formally defined as:
\begin{equation}
\texttt{INSTALL\_CMD\_LINE} \rightarrow \texttt{pip install} \; \texttt{PACKAGE\_NAMES} \; \backslash\texttt{n}.
\end{equation}
Crucially, since LLMs frequently recommend multiple packages in a single command, the parser must support iterative validation. Once an installation command is recognized, \name enforces intervention for every subsequent token sequence identified as a package name:
\begin{equation}
\texttt{PACKAGE\_NAMES} \rightarrow \texttt{PACKAGE\_NAME} \; (\texttt{PACKAGE\_NAME})^{*}.
\end{equation}

During inference, the context-aware parser operates in a continuous monitoring mode. As long as the output state corresponds to \texttt{NATURAL\_LANGUAGE} or \texttt{ORDINARY\_CODE\_LINE}, the package-name intervenor remains dormant, incurring zero interference. However, the moment the parser detects a state transition into \texttt{PACKAGE\_NAMES}, it immediately triggers active intervention. This design effectively shifts \name from passive context-tracking to strict validity enforcement, ensuring zero package hallucinations without compromising the generation of benign text.

\subsection{How to Intervene in LLM Generation?}
\label{3.3}

Once activated, the \textbf{package-name intervenor} assumes strict control over the generation process. Unlike post-hoc verification methods, this module operates synchronously during the decoding time. It functions in two coordinated stages: (i) it rigorously defines the legal generation space using a Deterministic Finite Automaton (DFA); and (ii) it performs intervention on the probability distribution via logits masking to render hallucinated packages theoretically unreachable.

\subsubsection{Legal State Verification via DFA}
To guarantee that every generated token contributes to a valid package name, \name compiles the authoritative package list into a DFA. 

\paragraph{DFA Construction.}
We formalize the package validity specification as a DFA $D=(Q, \Sigma, \delta, s_0, F)$, where $Q$ is the state space, $\Sigma$ is the character alphabet, $\delta: Q \times \Sigma \to Q$ is the deterministic transition function, $s_0 \in Q$ is the initial state, and $F \subseteq Q$ is the set of accepting states.
By treating each package name $p_i$ in the authoritative package list $P=\{p_1,\dots,p_N\}$ as a terminal string, the DFA serves as a deterministic realization of the following production rule:
\begin{equation}
\texttt{PACKAGE\_NAME} \rightarrow p_1 \mid p_2 \mid \dots \mid p_N.
\end{equation}
During inference, \name keeps track of the DFA state induced by the package-name prefix generated so far. When the model outputs the next token, \name interprets it as a character string $c_t$ and updates the state by consuming these characters sequentially:
\begin{equation}
s_t = \delta^{*}(s_{t-1}, c_t).
\end{equation}
where $\delta^{*}$ denotes the extended transition function over a character sequence.

\paragraph{Token Trie Integration.} The necessity of integrating a Token Trie arises fundamentally from the granularity mismatch between the generation unit of LLMs and the verification unit of the DFA. LLMs operate on a vocabulary $V$ of subword tokens, generating a multi-character string $v$ at each step. In contrast, the constructed DFA defines validity via character-level transitions $\delta: Q \times \Sigma \to Q$. Consequently, a single generation step by the LLM corresponds to a sequence of state transitions in the DFA, making direct verification impossible without decomposing the token.

To bridge this gap, we pre-build a \textbf{Token Trie} from the tokenizer's vocabulary. This Trie structurally maps each token $v \in V$ to its constituent character sequence. By synchronizing the traversal of the Token Trie with the DFA, we can determine whether the character sequence of a candidate token $v$ forms a valid path starting from the current DFA state $s_t$. Formally, a token $v$ is included in the feasible set $A(s_t)$ if and only if its character sequence allows the DFA to transition from $s_t$ to a valid subsequent state (i.e., not the empty set).

\subsubsection{Logits-Level Masking}
Once the set of feasible tokens $A(s_t)$ is identified, \name{} (which is denoted as $P$ below) enforces compliance at the probabilistic level. Specifically, we apply a logits mask that strictly forbids the generation of any token outside the legal set.

In standard autoregressive decoding, the LLM $L_\theta$ computes the next-token probability based on logits $\ell_t$:
\begin{equation}
L_\theta(x_t \mid x_{<t}) = \mathrm{softmax}(\ell_t).
\end{equation}
We model the intervened system as a reshaped distribution $L_\theta \parallel P$, where the original logits are modulated by a dynamic mask $m_t$ derived from $A(s_t)$:
\begin{equation}
(L_\theta \parallel P)(x_t \mid x_{<t}) = \mathrm{softmax}(\ell_t \oplus m_t).
\end{equation}
The mask $m_t$ acts as a binary gatekeeper over the vocabulary $V$:
\begin{equation}
m_t(v) = 
\begin{cases} 
1, & v \in A(s_t),\\
0, & v \notin A(s_t).
\end{cases}
\end{equation}
The operator $\oplus$ applies a mask to the logits:
\begin{equation}
(\ell_t \oplus m_t)[v] = 
\begin{cases} 
\ell_t[v], & \text{if } m_t(v)=1,\\
-\infty, & \text{if } m_t(v)=0.
\end{cases}
\end{equation}
By reshaping the distribution $L_\theta$, our proposed package-name intervenor ensures that the probability of sampling an invalid token becomes exactly zero. Consequently, every generated prefix is guaranteed to remain on some valid DFA path; if decoding terminates in an accepting state, the produced package name is guaranteed to be a valid entry in the authoritative package list.

\subsection{How to Efficiently Perform Monitoring?}
\label{3.4}

In practical scenarios, the DFA constructed from large-scale package lists exhibits an extremely large state space, leading to substantial overhead in each construction. Our extensive experimental results indicate that constructing the DFA for a large-scale package list is the primary bottleneck of inference latency. For instance, with the PyPI package list containing approximately 700k package names, constructing the DFA takes several seconds, whereas performing token-level validation using the Token Trie requires only a few milliseconds or even microseconds.

To efficiently perform monitoring in practical settings, we introduce a \textbf{DFA-caching mechanism} that decouples one-time automaton construction from runtime inference. Our key insight is that, for a fixed package list, the structure of the DFA remains unchanged during inference. Leveraging this property, \name pre-constructs a DFA checkpoint during initialization and stores it in cache. For each generation request, \name loads the cached DFA directly, thereby avoiding repeated construction.

From a runtime perspective, this DFA-caching mechanism ensures that the expensive cost of initializing the DFA is incurred only once. This “build once, reuse many” design effectively decouples inference latency from the size of the package list, leaving the online overhead dominated by lightweight DFA state transitions and logits-level masking operations, both of which are computationally stable and negligible during inference. This mechanism enables \name to operate efficiently in million-scale software ecosystems.

\section{Experimental Setup}
This section describes our experimental setup, including the benchmarks, baseline methods, evaluation metrics, selected models, and implementation details. Under this setup, we conduct a series of experiments to answer the following Research Questions (RQs):
\begin{itemize}[leftmargin=16pt]
  \item \textbf{RQ1: How effective is \name in eliminating package hallucinations?} 
  
  This RQ evaluates whether \name can eliminate package hallucinations across different models and compares its performance against existing baselines.

  \item \textbf{RQ2: What is the efficiency overhead introduced by \name?} 
  
  This RQ measures the efficiency impact of \name by comparing the average generation speed (or latency) before and after integrating \name.

  \item \textbf{RQ3: Does \name degrade the general code generation capability of LLMs?} 
  
  This RQ examines whether \name affects the general code generation capability of LLMs.

  \item \textbf{RQ4: How much does DFA-caching mechanism improve inference efficiency and scalability?}
  
  This RQ is studied via an ablation analysis comparing cached versus non-cached settings, and further characterizing how inference latency scales with increasing package-list sizes.
\end{itemize}

\subsection{Benchmarks }
\label{4.1}
To evaluate package hallucinations in LLMs, we adopt two benchmarks in our experiments: the \textbf{HFuzzer} framework~\cite{HFuzzer} and an open-source coding dataset released by \citeauthor{weHaveAPackageForYou}~\cite{weHaveAPackageForYou}. 

\ding{182} \textbf{HFuzzer~\cite{HFuzzer}.} We adopt HFuzzer as our primary evaluation framework. HFuzzer is a fuzzing framework designed to systematically probe failure modes of LLMs through iterative test case generation. Compared to conventional fixed test cases, HFuzzer can continuously explore diverse generation patterns and effectively induce previously unseen hallucinated packages.
In the experiments, HFuzzer involves two cooperating models: a \textbf{tester model} for test case generation, and a \textbf{target model} responsible for code generation and package recommendation, serving as the model under evaluation. We select \textbf{DeepSeek-V3}~\cite{deepseekv3} as the tester model, as its strong reasoning and coding capabilities support effective and diverse code-relevant test case generation.

\ding{183} \textbf{Package4U~\cite{weHaveAPackageForYou}.} In addition to HFuzzer, we also evaluate our approach on a dataset released by \citeauthor{weHaveAPackageForYou}~\cite{weHaveAPackageForYou}. Following the title of the original paper, we refer to this dataset as \textbf{Package4U}. Package4U contains approximately 40,000 coding tasks generated by LLMs or collected from Stack Overflow, with around half involving Python programming. Each task consists of a natural language problem description and code-related context. We use this dataset in a subset of our experiments. Detailed experimental settings and results are reported in the corresponding sections.

\subsection{Models}
\label{4.2}
We evaluate our approach on a set of representative LLMs, as summarized in Table~\ref{tab:models}. Since \name is model-agnostic and can be applied to different LLMs without modifying their parameters, we follow the model selection of prior work, i.e., HFuzzer~\cite{HFuzzer}, to ensure a fair comparison. The selected models cover different training objectives, including code-oriented and general-purpose LLMs, as well as different release periods.
\begin{itemize} [leftmargin=12pt]
    \item DeepSeek-Coder-Instruct-6.7B~\cite{DeepSeekCoder} is released by DeepSeek. It is an instruction-tuned, code-oriented LLM designed for programming tasks. In the following, we refer to this model as DeepSeek-Coder.
    \item CodeLlama-7b-Instruct-hf~\cite{codellama} is released by Meta. It is an instruction-tuned code model in the Code Llama family, widely adopted as a strong open-weight baseline for code generation. In the following, we refer to this model as CodeLlama.
    \item Qwen2.5-Coder-7B-Instruct~\cite{Qwen2.5-Coder} is released by the Qwen team (Alibaba). It is an instruction-tuned code model from the Qwen2.5 series, optimized for coding and tool-use style interactions. In the following, we refer to this model as Qwen2.5-Coder.
    \item Meta-Llama-3-8B-Instruct~\cite{llama3} is released by Meta. It is a general-purpose instruction-tuned LLM that provides a strong non-code-specialized baseline for evaluating robustness beyond code-centric training. In the following, we refer to this model as Meta-Llama-3.
    \item Meta-Llama-3.1-8B-Instruct~\cite{llama3.1} is released by Meta as an updated general-purpose instruction-tuned LLM. Compared with Llama-3-8B-Instruct, it improves overall instruction-following and capability. In the following, we refer to this model as Meta-Llama-3.1.
\end{itemize}

\begin{table}[t]
\centering
\caption{The Details of Selected Models.}
\label{tab:models}
\small
\begin{tabular}{lccc}
\toprule
\textbf{Model} & \textbf{Size} & \textbf{Release Date}  & \textbf{Download Count} \\
\midrule
DeepSeek-Coder-Instruct-6.7B~\cite{DeepSeekCoder} & 6.7B & 2023-11 & 56K \\ 
CodeLlama-7b-Instruct-hf~\cite{codellama}         & 7B   & 2023-08 & 19K \\
Qwen2.5-Coder-7B-Instruct~\cite{Qwen2.5-Coder}    & 7B   & 2024-09 & 1.18M \\
Meta-Llama-3-8B-Instruct~\cite{llama3}            & 8B   & 2024-04 & 1.53M \\
Meta-Llama-3.1-8B-Instruct~\cite{llama3.1}        & 8B   & 2024-07 & 10M \\
\bottomrule
\end{tabular}
\end{table}

\subsection{Baselines}
\label{4.3}
Following prior work on mitigating package hallucinations~\cite{weHaveAPackageForYou}, we include three representative baselines in our evaluation: \textbf{Retrieval-Augmented Generation (RAG)}, \textbf{Self-Refinement (SR)}, and \textbf{Supervised Fine-Tuning (SFT)}. Their details are introduced in Section~\ref{2.2}. we additionally include a stronger baseline, \textbf{RAG+SR}, which combines Retrieval-Augmented Generation with post-hoc Self-Refinement: we first apply RAG to inject retrieved package information into the prompt, and then run SR to verify and refine the model's generated installation commands. 

\subsection{Evaluation Metrics}
\label{4.4}
To evaluate the effectiveness of \name and the compared baselines, we adopt three metrics proposed by prior work~\cite{HFuzzer, weHaveAPackageForYou} that quantify hallucinated packages in model outputs, as described below:
\begin{itemize}[leftmargin=16pt]
    \item \textbf{Package Hallucination Rate (PHR)}. This metric is defined as the ratio of hallucinated packages to the total number of recommended packages. 
    \item \textbf{Response Hallucination Rate (RHR)}. This metric measures the percentage of model responses that contain at least one hallucinated package. 
    \item \textbf{Number of Unique Hallucinated Packages ($P_{hall-unique}$)}. This metric is analogous to a commonly used indicator in prior work~\cite{fuzzing}, i.e., the number of unique bugs, and reflects the diversity of hallucinated packages generated by the model.
\end{itemize}

\subsection{Implementation Details}
\label{4.5}
For all experiments, we use the PyPI \cite{pypi} index as the authoritative package list for DFA construction. This list comprises 706,618 valid package names at the time of our experiments, serving as the source for monitoring package-name generation.

For each target model listed in Table~\ref{tab:models}, we execute one full round of HFuzzer testing (with each round consisting of 1,000 iterations) with \name and all baselines. To ensure the determinism of experimental results, the temperature parameter of the tester model is fixed at 0, while that of the target model is set to 0.7 to strike a balance between generation creativity and output stability \cite{llm_settings}.

All experiments are conducted on a server equipped with eight NVIDIA A100 GPUs with 80 GB memory each, two AMD EPYC 7543 CPUs with 32 cores per socket (64 physical cores in total), 1 TB system memory, running Ubuntu 22.04 LTS.

\section{ Experimental Results}
\subsection{RQ1: How Effective is \name in Eliminating Package Hallucinations?}
\label{RQ1}
\begin{table}[t]
\centering
\caption{Comparative performance of \name{} against training-free baselines on the HFuzzer benchmark. $P_{total}$ denotes the total number of recommended packages, while $P_{hall}$ and $P_{hall-unique}$ represent the total and unique hallucinated packages, respectively. PHR and RHR indicate the package and response hallucination rates.}
\label{tab:results}
\small
\begin{tabularx}{\linewidth}{l l *{5}{>{\centering\arraybackslash}X}}
\toprule
\textbf{Model} & \textbf{Configuration}
& \textbf{$P_{\textit{total}}$}
& \textbf{$P_{\textit{hall}}$}
& \textbf{$P_{\textit{hall-uniq}}$}
& \textbf{PHR}
& \textbf{RHR} \\
\midrule
\multirow{5}{*}{DeepSeek-Coder}
 & vanilla & 1847 & 155 & 74 & 8.39\% & 11.60\% \\
 & RAG & 2288 & 132 & 87 & 5.77\% & 10.10\% \\
 & SR & 1496 & 113 & 47 & 7.55\% & 8.20\% \\
 & RAG + SR & 1750 & 108 & 46 & 6.17\% & 8.90\% \\
 \rowcolor{gray!19}
 & \textbf{\name{}} & \textbf{1648} & \textbf{0} & \textbf{0} & \textbf{0}\% & \textbf{0}\% \\
\midrule
\multirow{5}{*}{CodeLlama}
 & vanilla & 1004 & 63 & 63 & 6.27\% & 2.70\% \\
 & RAG & 690 & 41 & 39 & 5.94\% & 2.70\% \\
 & SR & 502 & 20 & 20 & 3.98\% & 1.70\% \\
 & RAG + SR & 231 & 3 & 3 & 1.30\% & 0.30\% \\
 \rowcolor{gray!19}
 & \textbf{\name{}} & \textbf{1514} & \textbf{0} & \textbf{0} & \textbf{0\%} & \textbf{0\%} \\
\midrule
\multirow{5}{*}{Qwen2.5-Coder}
 & vanilla & 553 & 18 & 11 & 3.25\% & 1.80\% \\
 & RAG & 600 & 20 & 13 & 3.33\% & 1.70\% \\
 & SR & 513 & 10 & 8 & 1.95\% & 0.90\% \\
 & RAG + SR & 551 & 14 & 13 & 2.54\% & 1.40\% \\
 \rowcolor{gray!19}
 & \textbf{\name{}} & \textbf{529} & \textbf{0} & \textbf{0} & \textbf{0}\% & \textbf{0}\% \\
\midrule
\multirow{5}{*}{Meta-Llama-3}
 & vanilla & 1623 & 80 & 38 & 4.93\% & 6.40\% \\
 & RAG & 1773 & 85 & 56 & 4.79\% & 7.10\% \\
 & SR & 1604 & 63 & 39 & 3.93\% & 5.10\% \\
 & RAG + SR & 1775 & 77 & 50 & 4.34\% & 6.20\% \\
 \rowcolor{gray!19}
 & \textbf{\name{}} & \textbf{1526} & \textbf{0} & \textbf{0} & \textbf{0}\% & \textbf{0}\% \\
\midrule
\multirow{5}{*}{Meta-Llama-3.1}
 & vanilla & 1979 & 113 & 88 & 5.71\% & 8.20\% \\
 & RAG & 2116 & 109 & 75 & 5.15\% & 9.00\% \\
 & SR & 1848 & 78 & 51 & 4.22\% & 7.00\% \\
 & RAG + SR & 2099 & 100 & 74 & 4.76\% & 8.90\% \\
 \rowcolor{gray!19}
 & \textbf{\name{}} & \textbf{1938} & \textbf{0} & \textbf{0} & \textbf{0}\% & \textbf{0}\% \\
\bottomrule 
\end{tabularx}
\end{table}

In this RQ, we compare \name with representative training-free and training-based approaches for mitigating package hallucinations.

\vspace{4pt}
\subsubsection{Comparison with training-free methods.} ~

\textbf{Settings.} We evaluate \name and the training-free baselines introduced in Section~\ref{4.3} on the five LLMs described in Section~\ref{4.2}, using HFuzzer. We report the main metrics---PHR, RHR, and $P_{\textit{hall-unique}}$---as defined in Section~\ref{4.4}. In addition, we include two auxiliary statistics: $P_{\textit{total}}$, the total number of packages recommended in each HFuzzer round, and $P_{\textit{hall}}$, the total number of hallucinated packages in each HFuzzer round.

\vspace{4pt}
\textbf{Results.} The results of different approaches are reported in Table ~\ref{tab:results}. 

\textbf{\name consistently outperforms all training-free baselines across every model, reducing all hallucination-related metrics to absolute zero.}
As shown in Table~\ref{tab:results}, \name eliminates package hallucinations on all five LLMs: for example, on DeepSeek-Coder, the vanilla setting exhibits a PHR/RHR of 8.39\%/11.60\%, while \name achieves 0\% PHR/RHR. Similar results hold for Meta-Llama-3.1 and Qwen2.5-Coder, where \name consistently drives all hallucination metrics to zero. These results indicate that \name provides a strict guarantee of eliminating package hallucinations and generalizes across model families.
In contrast, existing mitigation methods exhibit limited and sometimes unstable effectiveness. RAG can even \emph{increase} hallucination rates in several settings---for instance, on Qwen2.5-Coder, PHR increases from 3.25\% (vanilla) to 3.33\% (RAG), and on Meta-Llama-3, RHR increases from 6.40\% to 7.10\%---which we hypothesize may be due to prompt-injected information inducing overconfidence and encouraging unsupported outputs. SR yields more consistent reductions, but its hallucination rates remain strictly non-zero. For example, on DeepSeek-Coder it still leaves 113 hallucinated packages (PHR 7.55\%, RHR 8.20\%).

\textbf{\name does not significantly affect the number of recommended packages.}
Since \name operates by pruning all generation paths that would lead to hallucinated package names, a natural concern is that it may substantially reduce the number of package recommendations produced by the model. To examine this, we report $P_{\textit{total}}$, the total number of recommended packages in each fuzzing round. We find that across all models, the $P_{\textit{total}}$ achieved by \name is comparable to that of the vanilla model. This indicates that \name does not eliminate hallucinations by suppressing package recommendations; instead, it removes hallucinated packages while preserving the model’s tendency to recommend valid packages.
 
\vspace{4pt}
\subsubsection{Comparison with training-based method.} ~

\textbf{Settings.} \citeauthor{weHaveAPackageForYou}~\cite{weHaveAPackageForYou} mitigate package hallucinations by performing SFT on CodeLlama and DeepSeek-Coder using hallucination-free data. To faithfully reproduce their setting, we directly use the fine-tuned checkpoints released by the authors for evaluation. However, we observe that these SFT models exhibit noticeably degraded instruction-following behavior and often fail to consistently comply with the strict output format required by HFuzzer, making them incompatible with the HFuzzer evaluation pipeline.
Therefore, following \citeauthor{weHaveAPackageForYou}~\cite{weHaveAPackageForYou}, we evaluate training-based methods on the Package4U benchmark and report the corresponding package hallucination rates.

\vspace{4pt}
\textbf{Results.} The results are shown in Table~\ref{tab:sft}.

\textbf{Despite requiring no additional training or parameter updates, \name offers a fundamentally stronger guarantee of package-hallucination elimination than training-based approaches.}
As shown in Table~\ref{tab:sft}, while SFT substantially reduces hallucination rates, it does not eliminate them: non-zero hallucination rates of 2.66\% on DeepSeek-Coder and 10.27\% on CodeLlama remain, which still pose non-negligible security risks.
In contrast, \name consistently reduces the package hallucination rate to zero across all experimental configurations. These results demonstrate that, even compared with training-based methods that require costly fine-tuning pipelines, \name not only avoids training overhead but also provides a more robust and deterministic solution for hallucination elimination.

\begin{tcolorbox}[
    enhanced,                          
    boxrule=0pt,    
    frame hidden,
    colback=gray!12,                                                    
    arc=5pt,                             
    shadow={1.5pt}{-1.5pt}{0pt}{gray!50},
    top=6pt, bottom=6pt, left=8pt, right=8pt, 
    before skip=10pt, after skip=10pt  
]
\textbf{Answer to RQ1.}
\name can stably eliminate package hallucinations, whereas existing mitigation approaches can only reduce package hallucinations and cannot eradicate them.
\end{tcolorbox}

\begin{table}[t]
\centering
\caption{Package Hallucination Rates (PHR) comparison between \name{} and the SFT baseline.}
\label{tab:sft}
\small
\begin{tabular}{lcc}
\toprule
 & \textbf{DeepSeek-Coder} & \textbf{CodeLlama} \\
\midrule
Vanilla               & 16.14\% & 26.28\% \\
SFT                             & 2.66\% & 10.27\% \\
\rowcolor{gray!18}\textbf{\name{}}                         & \textbf{0\%} & \textbf{0\%} \\
\bottomrule
\end{tabular}
\end{table}

\subsection{RQ2: What is the Efficiency Overhead Introduced by \name?}
\label{RQ2}

Although \name eliminates package hallucinations, it must remain lightweight to be practical in real deployments. Therefore, we quantify the overhead introduced by \name and compare it against existing approaches.

\definecolor{MyBlue}{RGB}{65, 113, 156}
\begin{table}[t!]
    \centering
    \renewcommand{\arraystretch}{1.2}
    \caption{Average generation time (seconds) across different LLMs. 
    For each model, the best result is highlighted in bold. Numbers in parentheses indicate the relative inference time normalized by the corresponding vanilla model.}
    \label{tab:RQ2}
    \vspace{2mm}
    \resizebox{0.9\textwidth}{!}{
    \begin{tabular}{lccccc}
        \toprule
        \textbf{Method} & \textbf{DeepSeek-Coder} & \textbf{CodeLlama} & \textbf{Qwen2.5-Coder} & \textbf{Llama-3} & \textbf{Llama-3.1} \\
        \midrule
        Vanilla    & 1.9809 & 2.4296 & 0.8019 & 0.9020 & 1.5277 \\
        \hline
        RAG    & 2.4751 \textcolor{gray}{($\times$1.25)} & 3.9064 \textcolor{gray}{($\times$1.61)} & 0.8797 \textcolor{gray}{($\times$1.10)} & 1.5359 \textcolor{gray}{($\times$1.70)} & 1.8978 \textcolor{gray}{($\times$1.24)} \\
        SR     & 3.8439 \textcolor{gray}{($\times$1.94)} & 3.3827 \textcolor{gray}{($\times$1.39)} & 1.4026 \textcolor{gray}{($\times$1.75)} & \textbf{1.0031} \textcolor{gray}{($\times$1.11)} & \textbf{1.6115} \textcolor{gray}{($\times$1.05)} \\
        RAG+SR & 5.2509 \textcolor{gray}{($\times$2.65)} & 4.4491 \textcolor{gray}{($\times$1.83)} & 1.0595 \textcolor{gray}{($\times$1.32)} & 1.3596 \textcolor{gray}{($\times$1.51)} & 2.1525 \textcolor{gray}{($\times$1.41)} \\
        \rowcolor{gray!18} \textbf{Ours} & \textbf{2.2351} \textcolor{MyBlue}{\textbf{($\times$1.13)}} & \textbf{3.0130} \textcolor{MyBlue}{\textbf{($\times$1.24)}} & \textbf{0.8596} \textcolor{MyBlue}{\textbf{($\times$1.07)}} & 1.1515 \textcolor{MyBlue}{\textbf{($\times$1.28)}} & 1.6464 \textcolor{MyBlue}{\textbf{($\times$1.08)}} \\
        \bottomrule
    \end{tabular}
    }
\end{table}

\vspace{4pt}
\textbf{Settings.} We measure efficiency on HFuzzer. For each HFuzzer round, we evaluate the five models introduced in Section~\ref{4.2} with \name and all the training-free baselines introduced in Section~\ref{4.3}, and report the average generation time (in seconds).

\begin{figure}[t!]
    \vspace{5pt}
    \centering
    \includegraphics[width=0.62\textwidth]{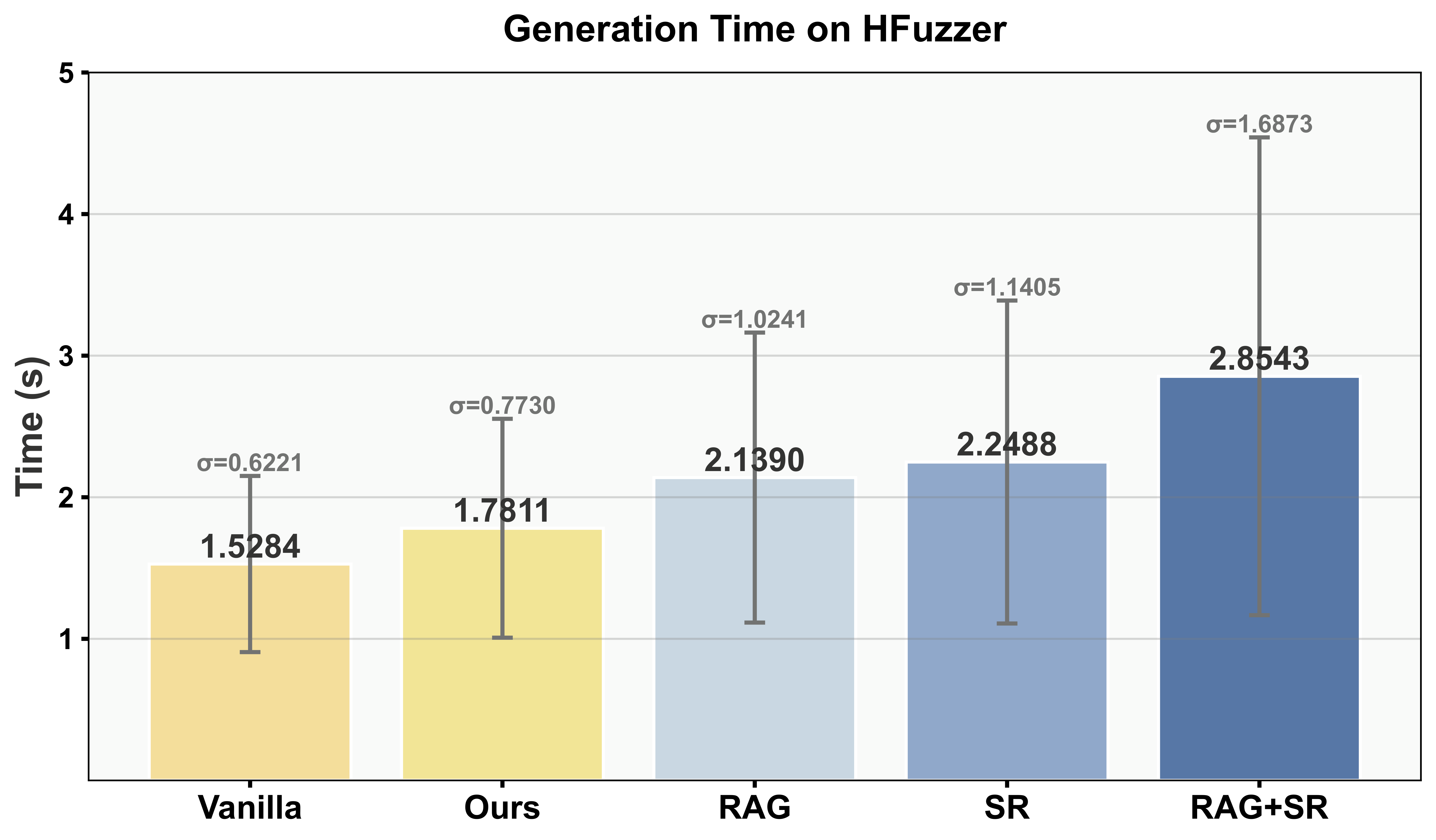}
    \caption{Average generation time and standard deviation of different methods on HFuzzer, aggregated across LLMs.}
    \label{fig:gen_speed}
\end{figure}

\vspace{4pt}
\textbf{Results.} The results are summarized in Table~\ref{tab:RQ2} and visualized in Figure~\ref{fig:gen_speed}.

\textbf{The average overhead is small.} Table~\ref{tab:RQ2} shows that integrating \name introduces only a modest and consistent slowdown across all five models, with a normalized inference-time ratio between \textbf{$\times$1.07 and $\times$1.28}. In absolute terms, the added latency is limited: +0.0577s on Qwen2.5-Coder (0.8019s $\rightarrow$ 0.8596s, $\times$1.07), +0.1187s on Llama-3.1 (1.5277s $\rightarrow$ 1.6464s, $\times$1.08), +0.2495s on Llama-3 (0.9020s $\rightarrow$ 1.1515s, $\times$1.28), +0.2542s on DeepSeek-Coder (1.9809s $\rightarrow$ 2.2351s, $\times$1.13), and $\times$1.24 on CodeLlama (2.4296s $\rightarrow$ 3.0130s). Although the relative overhead can reach $\times$1.28, the absolute latency increase is small, around 0.2s in the worst case. Moreover, \name is among the most time-efficient methods overall: it achieves the best latency on DeepSeek-Coder, Qwen2.5-Coder, and CodeLlama, and the second-best latency on Llama-3.1 and Llama-3. In contrast, RAG and SR often incur substantially higher overhead, and RAG+SR can be particularly expensive (e.g., up to $\times$2.65 on DeepSeek-Coder).

\textbf{The latency variance is small and stable.} Figure~\ref{fig:gen_speed} further reports the average generation time and standard deviation aggregated across models. Compared with SR and RAG+SR, \name exhibits noticeably smaller variance, indicating that its overhead is more predictable and does not suffer from occasional latency spikes. This stability is expected because \name performs only lightweight per-step operations (DFA state transitions and logits masking) during decoding, whereas SR and RAG+SR may introduce variable amounts of additional computation due to refinement iterations (and RAG also adds retrieval-related variability).

\begin{tcolorbox}[
    enhanced,
    boxrule=0pt,
    frame hidden,
    colback=gray!12,
    arc=5pt,
    shadow={1.5pt}{-1.5pt}{0pt}{gray!50},
    top=6pt, bottom=6pt, left=8pt, right=8pt,
    before skip=10pt, after skip=10pt
]
\textbf{Answer to RQ2.}
On HFuzzer, \name incurs only a small and stable efficiency overhead, whereas existing training-free baselines introduce noticeably higher and more variable time costs.
\end{tcolorbox}

\subsection{RQ3: Does \name Degrade the General Code Generation Capability of LLMs?}
\label{5.3}

It is crucial to verify that our proposed \name does not compromise the model's general code generation capabilities. To this end, we evaluate whether \name affects functional correctness on a standard code generation benchmark.

\definecolor{MyBlue}{RGB}{65, 113, 156}
\definecolor{MyRed}{RGB}{150,0,0}

\begin{table}[t]
\centering
\caption{Comparison of code generation quality (Pass@1) on HumanEval. Values in parentheses represent the performance drop relative to the vanilla model.}
\label{tab:humaneval}
\small
\begin{tabular}{lccc}
\toprule
\textbf{Model} & \textbf{Vanilla} & \textbf{With \name{}} & \textbf{With SFT} \\
\midrule
DeepSeek-Coder-6.7B & 72.6\% & 72.6\% \textcolor{MyBlue}{($\downarrow$ 0\%)}& 45.7\% \textcolor{MyRed}{($\downarrow$ 26.9\%)} \\
CodeLlama-7B & 18.3\% & 18.3\% \textcolor{MyBlue}{($\downarrow$ 0\%)} & 4.3\% \textcolor{MyRed}{($\downarrow$ 14.0\%)}\\
Qwen2.5-Coder-7B & 79.3\% & 79.3\% \textcolor{MyBlue}{($\downarrow$ 0\%)} & - \\
Meta-Llama-3-8B & 53.0\% & 53.0\% \textcolor{MyBlue}{($\downarrow$ 0\%)}& - \\
Meta-Llama-3.1-8B & 60.4\% & 60.4\% \textcolor{MyBlue}{($\downarrow$ 0\%)}& - \\
\bottomrule
\end{tabular}
\end{table}

\vspace{4pt}
\textbf{Settings.} Following the HumanEval~\cite{humaneval} protocol adopted by \citeauthor{weHaveAPackageForYou}~\cite{weHaveAPackageForYou}, we evaluate the same models before and after integrating \name, and report Pass@1 as the metric for functional correctness. For stability, we set the decoding temperature to 0 for all runs in this RQ.

\vspace{4pt}
\textbf{Results.} The results are shown in Table~\ref{tab:humaneval}.

\textbf{Across all evaluated models, HumanEval performance remains unchanged after incorporating \name.}
As shown in Table~\ref{tab:humaneval}, the Pass@1 scores remain unchanged with and without \name, indicating that our approach does not degrade general code generation capability. We attribute this to the fact that \name only intervenes when the model is generating package names following installation commands. As a result, \name eliminates hallucinated packages without interfering with ordinary code generation, thereby preserving the original utility.

\textbf{The utility preservation of \name contrasts sharply with SFT.} Prior work reports that although SFT can significantly reduce package hallucinations, it may noticeably hurt general code generation performance. On HumanEval, the Pass@1 scores of DeepSeek-Coder-6.7B and CodeLlama-7B drop by 26.9\% and 14.0\%, respectively~\cite{weHaveAPackageForYou}. This degradation is expected because SFT changes model weights and may inadvertently harm general code generation performance.

\begin{tcolorbox}[
    enhanced,                          
    boxrule=0pt,    
    frame hidden,
    colback=gray!12,                                                    
    arc=5pt,                             
    shadow={1.5pt}{-1.5pt}{0pt}{gray!50},
    top=6pt, bottom=6pt, left=8pt, right=8pt, 
    before skip=10pt, after skip=10pt   
]
\textbf{Answer to RQ3.}
On HumanEval, integrating \name does not change Pass@1 across all evaluated models, indicating that \name preserves general code generation capability.
\end{tcolorbox}

\subsection{RQ4: How Much does DFA-Caching Mechanism Improve Inference Efficiency and Scalability?}
In this RQ, we conduct an ablation study to evaluate the effectiveness of the DFA-caching mechanism and to examine how decoding latency scales with increasing package list sizes.

\definecolor{MyBlue}{RGB}{65, 113, 156}
\definecolor{MyRed}{RGB}{150,0,0}

\begin{table}[!t]
\caption{Comparison of average generation time on HFuzzer with and without caching across LLMs. Values in parentheses represent the time multiplier relative to the vanilla model.}
\label{tab: cache}
\centering
\small
\renewcommand{\arraystretch}{1.1}
\setlength{\tabcolsep}{8pt} 
\begin{tabular}{lc}
\specialrule{1pt}{0.7pt}{0pt} 
\textbf{Model} & \textbf{Average Generation Time (s)} \\
\midrule
\textbf{DeepSeek-Coder-6.7B} & 1.9809 \\
\rowcolor{gray!8} \quad + \name{} (w/o cache) & 5.9150 \textcolor{MyRed}{($\times$2.99)} \\
\rowcolor{gray!19} \quad + \name{} & 2.2351 \textcolor{MyBlue}{($\times$1.13)} \\ 
\textbf{CodeLlama-7B} & 2.4296 \\
\rowcolor{gray!8} \quad + \name{} (w/o cache) & 6.8237 \textcolor{MyRed}{($\times$2.81)} \\
\rowcolor{gray!19} \quad + \name{} & 3.0130 \textcolor{MyBlue}{($\times$1.24)} \\
\textbf{Qwen2.5-Coder-7B} & 0.8019 \\
\rowcolor{gray!8} \quad + \name{} (w/o cache) & 5.0078 \textcolor{MyRed}{($\times$6.24)} \\
\rowcolor{gray!19} \quad + \name{} & 0.8596 \textcolor{MyBlue}{($\times$1.07)} \\
\textbf{Meta-Llama-3-8B} & 0.9020 \\
\rowcolor{gray!8} \quad + \name{} (w/o cache) & 5.3106 \textcolor{MyRed}{($\times$5.89)} \\
\rowcolor{gray!19} \quad + \name{} & 1.1515 \textcolor{MyBlue}{($\times$1.28)} \\
\textbf{Meta-Llama-3.1-8B} & 1.5277 \\
\rowcolor{gray!8} \quad + \name{} (w/o cache) & 6.1537 \textcolor{MyRed}{($\times$4.03)} \\
\rowcolor{gray!19} \quad + \name{} & 1.6464 \textcolor{MyBlue}{($\times$1.08)} \\
\specialrule{1pt}{0.7pt}{0pt} 
\end{tabular}
\end{table}

\vspace{4pt}
\subsubsection{Impact of DFA-Caching on Inference Latency} ~

\textbf{Settings.} We evaluate the inference time of five LLMs introduced in Section~\ref{4.2} under the HFuzzer benchmark across three configurations: the original model, \name integrated without caching (w/o cache), and the full \name with caching enabled. 

\vspace{4pt}
\textbf{Results.} Experimental results are demonstrated in Table ~\ref{tab: cache}.

\textbf{Overall, enabling DFA-caching substantially improves decoding efficiency, bringing inference time close to that of the original model.}
As shown in Table~\ref{tab: cache}, disabling caching incurs a large and consistent overhead across all models (typically \textbf{$\times$2.8--$\times$6.2} relative to vanilla), because the DFA must be repeatedly constructed and initialized during online inference. With caching enabled, latency drops sharply and remains close to the vanilla baseline, reducing the overhead to only \textbf{$\times$1.07--$\times$1.28}. This gap is particularly pronounced for faster base models (e.g., Qwen2.5-Coder and Llama-3), where on-the-fly DFA construction dominates end-to-end decoding time. Overall, DFA-caching removes the one-time, high-cost automaton construction from the online path, leaving \name’s runtime overhead dominated by lightweight DFA state transitions and logits masking.

\begin{table}[t]
\centering
\caption{Efficiency gains from parser caching: comparison between one-time offline construction and rapid runtime loading across multi-million package scales.}
\label{tab:scaling}
\small
\begin{tabular}{lcccc}
\toprule
\textbf{Time Cost} & \textbf{3000k} & \textbf{700k} & \textbf{70k} & \textbf{7k} \\
\midrule
DFA Construction & 21.121s & 4.407s & 0.265s & 0.021s \\
\rowcolor{gray!19}DFA Loading      & 0.247s & 0.083s & 0.020s & 0.002s \\
\bottomrule
\end{tabular}
\vspace{-0.1in}
\end{table}

\vspace{4pt}
\subsubsection{Efficient Scaling to Million-Level Package List Size} ~

We further evaluate the scalability of \name by measuring the time overhead of two critical stages in the initialization of the monitoring mechanism: DFA construction and DFA loading.

As shown in Table \ref{tab:scaling}, DFA construction exhibits strong sensitivity to scale. When the package list grows from 7,000 to 3,000,000 entries (approximately a 428× increase), the construction time increases starkly from 0.021s to 21.121s (about a 1000× slowdown). This behavior stems from the rapid expansion of the DFA state space induced by millions of distinct package names. In contrast, DFA loading time is largely insensitive to scale, remaining as low as 0.247s even at the 3,000k scale. This disparity clearly demonstrates the necessity of DFA-caching mechanism, which converts the substantial compilation cost into a one-time offline overhead.

Without the DFA-caching mechanism, \name would be required to reconstruct the DFA online for every inference request. At a future scale where the package list reaches 3,000k entries, this would incur a construction cost of up to 21.121s per inference—an overhead that is clearly unacceptable for latency-sensitive LLM inference workloads. By introducing DFA-caching, we shift the expensive DFA construction entirely to an offline phase, leaving only a lightweight loading operation at runtime. Empirically, even at the 3,000k scale, DFA loading takes only 0.247s, achieving an approximately 85× speedup compared to construction time costs. This “build once, reuse many” strategy effectively decouples inference latency from package list scale, enabling \name to support million-scale, continuously evolving ecosystems such as PyPI without incurring perceptible overhead during online inference.

\begin{tcolorbox}[
    enhanced,                          
    boxrule=0pt,    
    frame hidden,
    colback=gray!12,                                                    
    arc=5pt,                             
    shadow={1.5pt}{-1.5pt}{0pt}{gray!50},
    top=6pt, bottom=6pt, left=8pt, right=8pt, 
    before skip=10pt, after skip=10pt 
]
\textbf{Answer to RQ4.} Overall, these results demonstrate the effectiveness of the DFA caching mechanism in substantially reducing inference latency. Moreover, the consistently low DFA loading cost across large package lists confirms that \name scales to real-world, evolving software ecosystems.
\end{tcolorbox}

\section{Discussion}
In this section, we discuss several key aspects of \name{} beyond the quantitative results. We first analyze why \name{} is able to achieve zero package hallucination under our evaluated settings, and clarify the scope of this guarantee. We then examine its generalizability across package ecosystems, highlighting how the framework can be extended beyond the PyPI setting to support diverse installation commands and customizable package lists. Finally, we systematically analyze potential threats to the validity of our study.

\subsection{Understanding Why Zero Package Hallucination Is Achievable}
A central result of our evaluation is that \name{} consistently reduces package hallucination rates to zero across all tested models and settings. This outcome may appear unusually strong, especially given that prior work has shown hallucinations to be pervasive and difficult to eliminate. We discuss the key reasons why this guarantee is achievable in our setting.

First, package-name generation fundamentally differs from many other hallucination-prone generation tasks. Unlike open-ended natural language, valid package names form a finite and enumerable list. This property enables \name{} to derive an exact validity specification and to structurally exclude invalid package names from the generation space, rather than attempting to correct hallucinations after they occur.

Second, \name{} enforces validity directly at the decoding level, rather than relying on prompt engineering, self-verification, or model retraining. By applying logits-level masking based on an automaton constructed from an authoritative package list, \name{} guarantees that every generated package-name prefix remains consistent with at least one valid package entry. As a result, hallucinated packages are not merely discouraged but rendered unreachable by construction.

It is important to emphasize that this guarantee is conditional on the correctness and completeness of the given authoritative package list. Under these assumptions, \name{}  ensures that any generated package name is valid according to the repository at generation time.

\subsection{Generalization Across Package Ecosystems}
\label{6.2}
Our approach is not limited to the PyPI setting (i.e., monitoring package names only after \texttt{pip install}). Owing to its context-aware parser design, \name{} can be readily adapted, in principle, to different software ecosystems by applying ecosystem-specific package lists to installation commands such as \texttt{conda install} and \texttt{npm install}. In addition, the set of legitimate packages is not fixed and can be manually imported and customized to accommodate specific environments or private dependency scenarios. This endows \name{} with robust generalizability and practical deployment flexibility across various package ecosystems.

\subsection{Threats to Validity}
We now discuss potential threats to the validity of our study and how we mitigate them:

\ding{182} \textbf{Generalizability Across Models.} One potential threat concerns whether our findings generalize beyond the specific models. To reduce this threat, we conducted experiments on five widely used LLMs with diverse architectures, training data, and instruction-tuning strategies. In addition, we evaluated \name{} using HFuzzer, which generates a broad and adversarial set of dependency-related prompts. These choices increase confidence that our results are not artifacts of a particular model or prompt distribution. 

\ding{183} \textbf{Dependence on Authoritative Package Lists.} \name relies on an externally maintained authoritative package list as the ground truth for package-name intervenor. Any errors or omissions in this list (e.g., misrecorded entries or missing packages) could undermine the correctness of \name. In our experiments, we obtained the package list via PyPI’s official API. However, package ecosystems evolve continuously: new packages are published and existing ones may be deprecated over time. To maintain long-term effectiveness, a practical deployment can incorporate a periodic update mechanism to refresh the authoritative list snapshot and synchronize it with the latest ecosystem state, thereby preserving the hallucination-prevention guarantee.

\ding{184} \textbf{Impact of Deployment Environments.} For LLMs, generation speed is a critical metric. In our experiments, we evaluated a range of speed-related indicators (e.g., model generation latency, DFA construction time). These metrics typically fall within the millisecond-to-second range, making them potentially sensitive to variations in deployment environments (e.g., differences in GPU or CPU specifications). However, this threat is mitigated by the fact that our core effectiveness (e.g., hallucination-prevention) is independent of such environmental factors. To further support reproducibility across environments, we have made detailed experimental settings available in the Supplementary Materials, and our code repository is publicly accessible. These efforts allow our experiments to be replicated across different devices without compromising the validity of core conclusions.

\ding{185} \textbf{Scope of Evaluated Contexts and Downstream Executability.}
Our current implementation and evaluation focus on chat-based PyPI dependency recommendations after \texttt{pip install} commands. Therefore, our findings should be interpreted within this monitored setting, and do not yet cover broader dependency specifications such as \texttt{requirements.txt}, Dockerfiles, CI configurations, or end-to-end coding-agent workflows. In addition, \name targets
package-name validity rather than full program executability, since a valid package may still be API-incompatible, version-sensitive, or environment-dependent. We leave these broader evaluations to future work.

Overall, \name{} demonstrates that package hallucinations—often regarded as a challenging failure mode of LLMs—can be prevented under well-defined monitoring mechanism. By shifting from post-hoc mitigation to design-level prevention, our approach provides strong package-name validity guarantees guarantees while preserving generation quality and efficiency. At the same time, we acknowledge that these guarantees do not cover all dependency specification contexts or full program executability, which points to promising directions for future research on trustworthy software generation.

\section{Conclusion}

In this paper, we propose \name, a novel framework capable of rigorously eliminating package hallucinations in Large Language Models. Operating as a training-free and plug-and-play solution, \name leverages the deterministic nature of package validity to enforce strict compliance during the decoding process. Our extensive experiments demonstrate that \name consistently reduces package hallucination rates to absolute zero across all evaluated models. Crucially, it achieves this strict guarantee with negligible inference overhead and without compromising the models' general code generation capabilities. By closing the gap between generative probability and factual validity, we believe \name paves the way for the trustworthy adoption of LLM in real-world software development.

\section*{Data Availability}
The source code is provided at: \url{https://github.com/THU-Agent/PackMonitor}.

\section*{Acknowledgments}
This work was supported by Fundamental and Interdisciplinary Disciplines Breakthrough Plan of the Ministry of Education of China (No. JYB2025XDXM101), the Research Grant of Tsinghua-Tencent Joint Laboratory for Internet Innovation Technology.

This work was also supported by  Tsinghua University - Keystone Electrical (Zhejiang) Co.,Ltd Joint Research Center for Embodied Multimodal Artificial Intelligence (JCEMAI) and  the Beijing Natural Science Foundation under Grant No. 4264107.

\bibliographystyle{ACM-Reference-Format}
\bibliography{sample-base}

\end{document}